\documentclass[twoside]{article}

\begin{document}
\setcounter{equation}{0}
\begin{center}
\Large \bf  The Hamilton Operator and Quantum\\
            Vacuum for Nonconformal Scalar Fields\\
            in the Homogeneous and Isotropic Space
\end{center}
      \vspace{2mm}
\begin{center}
{\large Yu.V.\,Pavlov}\\[7mm]
   Institute of Mechanical Engineering, Russian Academy of Sciences,\\
      61 Bolshoy, V.O., St.Petersburg, 199178, Russia\\
                 e-mail: pavlov@ipme.ru
\end{center}

\begin{abstract}
\noindent   {\bf Abstract.} \
The diagonalization of the metrical and canonical Hamilton operators
of a scalar field with an arbitrary coupling, with a curvature in
$N$-dimensional homogeneous isotropic space is considered in this paper.
    The energy spectrum of the corresponding quasiparticles is obtained
and then the modified energy-momentum tensor is constructed;
the latter  coincides with the metrical energy-momentum tensor for
conformal scalar field.
    Under the diagonalization of corresponding Hamilton operator
the energies of relevant particles of a nonconformal field
are equal to the oscillator frequencies, and the density of such particles
created in a nonstationary metric is finite.
    It is shown that the modified Hamilton operator can be constructed
as a canonical Hamilton operator under the special choice of variables.

\end{abstract}

\section{Introduction}
    Our aim in this paper is the investigation of the Hamiltonian
diagonalization method and the definition of the Hamilton operator and
quantum vacuum of nonconformal scalar field in nonstationary homogeneous
isotropic space.
    Quantum field theory in curved space-time
(see monographs~\cite{GMM,BD}) has important applications to
cosmology and astrophysics.
  However there are several problems that have not been finally solved
until the present time.
    One of them is the definition of vacuum state and the notion of
elementary particle in curved space-time;
    this is due to the absence of the group of symmetries such as the
Poincare group in Minkowsky space.
    This problem for nonconformal scalar field is under active discussion
even in the case of homogeneous isotropic
space~\cite{Diag,Redmount99,Lindig}.
    As a consequence of various definitions of vacuum states we have
a~variety of calculated quantum characteristics of nonconformal scalar
fields in curved space.

    In~\cite{Pv,BLM} it was shown that in the case of arbitrary
coupling of scalar fields with curvature additional, nonconformal
contributions are dominant in vacuum averages of the energy-momentum tensor.
   It should be also mentioned that the investigation of nonconformal scalar
fields is not only of independent interest;
this investigation is caused by impossibility of preservation of conformal
invariation of effective action and the usual action in the
case of interacting quantized field~\cite{BD}.

    In the definition of vacuum and in the formulation of the problem
of particles creation in curved space-time two known approaches
are widely used: the Hamiltonian diagonalization procedure~\cite{GMM}
offered in~\cite{G69,G71}, and the so-called "adiabatic" procedure~\cite{BD}
offered in~\cite{Parker}.
    Supposing that a quantum of energy corresponds to a particle, then
observation of particles at some moment (according to quantum mechanics)
means to find the Hamilton operator's eigenstate;
   this is taken into account automatically in the diagonalization approach.
   In nonstationary metrics the diagonalization procedure is realized by
the time-dependent Bogolyubov transformations (see below).
   If the operators of these transformations are Hilbert-Schmidt
operators then the representations of commutation relations
are unitarily equivalent for both the old and the new
creation and annihilation operators~\cite{Berezin}.
   However, the use of the Hamilton operator  constructed from the metrical
energy-momentum tensor, successful in the conformal case~\cite{GMM},
leads to the difficulties related to an infinite density of created
particles in the nonconformal case~\cite{Fulling}.
    At the same time essential problems and ambiguities take place in
adiabatic approach also~\cite{Lindig}.

    In this paper we will consider the complex scalar field with arbitrary
coupling, with curvature in $N$-dimensional homogeneous isotropic space.
    In section~2 all necessary information is given and nonconformal scalar
field quantizing in $N$-dimensional homogeneous isotropic space is
defined.
    In section~3 the metrical and canonical Hamilton operators
diagonalization is carried out, and the energies of corresponding
quasiparticles are calculated, and conditions connected with demand of
the Hamilton operators diagonality are investigated.
    In section~4 the modified energy-momentum tensor is defined so that
the quasiparticles from diagonalization of corresponding Hamilton operator
have energies coinciding with the oscillator frequency of the wave equation.
    It is shown that such Hamiltonian can be defined as canonical
under a certain choice of canonical variables.
    It is proved that the density of particles being created in
a~nonstationary metric is finite and the results of given
investigations are summarize.

    The system of units in which the Planck constant ($\hbar $) and
light velocity are equal 1  is used in the paper.

\section{Quantizing of scalar field in homogeneous\\
         isotropic space}

    We consider the complex scalar field $\phi(x)$ of mass $m$ satisfying
the equation
\begin{equation}
 ({\nabla}_i {\nabla}^i + \xi R +m^2) \, \phi(x)=0 \, , \label{1}
\end{equation}
    where ${\nabla}_i$ is covariant differentiation,
$R$ is the scalar curvature, $x=(t,{\bf x})$, \
$\xi$~is the coupling constant.
    The value $ \xi =\xi_c =(N-2)/\,[\,4\,(N-1)] $
corresponds to conformal coupling in space-time of dimension $N$
($ \xi_c=1/6 $ if $N=4$).
    The equation (\ref{1}) is conformaly
invariant if $\xi=\xi_c $ and $m=0$;
   the value $\xi=0$  reduces to the case of minimal coupling.

   The metric of $N$-dimensional homogeneous isotropic space-time is
\begin{equation}
 ds^2=g_{ik}dx^i\,dx^k=dt^2-a^2(t)\,dl^2=a^2(\eta)\,(d{\eta}^2 - dl^2) \,,
 \label{2}
\end{equation}
    where $dl^2=\gamma_{\alpha \beta} \,dx^\alpha \,dx^\beta $
is the metric of ($N-1$)-dimensional space with the constant curvature
$K=0, \pm 1$.

    The equation (\ref{1}) can be obtain by varying the action with
Lagrangian density
\begin{equation}
L(x)=\sqrt{|g|}\ [\,g^{ik}\partial_i\phi^*\partial_k\phi -(m^2+\xi R)
\phi^* \phi \,] \, ,
\label{3}
\end{equation}
    where $g=\mbox{det}(g_{ik})$.

    The canonical energy-momentum tensor of the scalar field is
\begin{equation}
T_{ik}^{can}=\partial_i\phi^* \partial_k\phi+
\partial_k\phi^* \partial_i\phi-g_{ik} |g|^{-1/2}L(x) \, .
\label{4}
\end{equation}

     The metrical energy-momentum tensor which can be obtained by
varying the action of $g_{ik}$ has a form~\cite{ChT}:
\begin{equation}
T_{ik}=T_{ik}^{can} - 2 \, \xi \, [ R_{ik} + \nabla_i\nabla_k -
g_{ik}\nabla_j\nabla^j ] \, \phi^* \phi \, ,
\label{5}
\end{equation}
    where $R_{ik} $ is Ricci tensor.

    In the metric (\ref{2}) the equation (\ref{1}) takes the form
\begin{equation}
\phi''+(N-2) \left(\frac{a'}{a}\right) \phi'-\Delta_{N-1}\,\phi+
(m^2+\xi R)\,a^2\phi=0 \, ,
\label{6}
\end{equation}
    where $\Delta_{N-1}$  is the Laplace-Beltrami operator in
($N-1$)-dimensional space, and the prime denotes the derivative with
conformal time $\eta$.

    For the function $\tilde{\phi}= a^{(N-2)/\,2}\phi $
the equation (\ref{6}) takes the form without the first derivative in time
\begin{equation}
\tilde{\phi}'' -\Delta_{N-1}\,\tilde{\phi} +
\left(m^2 a^2-\Delta\xi a^2 R+ ((N-2)/\,2)^2 \,K \right)\,\tilde{\phi}=0 \,,
\label{7}
\end{equation}
    where  $ \Delta \xi=\xi_c-\xi $. \
The variables in the equations (\ref{6}), (\ref{7}) can be separated;
namely, for
$ \tilde{\phi}= g_\lambda (\eta) \Phi_J({\bf x}) $   we have
\begin{equation}
g_\lambda''(\eta)+\Omega^2(\eta)\,g_\lambda(\eta)=0 \,,
\label{8}
\end{equation}
    and
\begin{equation}
\Delta_{N-1}\,\Phi_J=-(\lambda^2-((N-2)/\,2)^2\,K )\,\Phi_J \,;
\label{9}
\end{equation}
     $\Omega(\eta) $ is the oscillator frequency
\begin{equation}
\Omega^2(\eta)=m^2 a^2 + \lambda^2 -\Delta \xi\,a^2 R \,,
\label{10}
\end{equation}
    $J$ is a set of indexes (quantum numbers) numbering the eigenfunctions
of the Laplace-Beltrami operator. It should be noted that the eigenvalues
of the operator $-\Delta_{N-1} $ are not negative and we have the
       inequality
$$
\lambda^2-((N-2)/\,2)^2\,K \ge 0 \,.
$$

    For quantization we decompose the field $ \tilde{\phi}(x) $
by the complete set of the solutions of~(\ref{7}), i.e.
\begin{equation}
\tilde{\phi}(x)=\int \! d\mu(J)\,\biggl[\,\tilde{\phi}^{(-)}_{\bar{J}}
\,a^{(-)}_{\bar{J}} + \tilde{\phi}^{(+)}_{J}\,a^{(+)}_{J}\,\biggr] \ ;
\label{11}
\end{equation}
    here $ d\mu (J) $ is the measure in the space of the Laplace-Beltrami
$ \Delta_{N-1} $ eigenvalues
\begin{equation}
\tilde{\phi}^{(+)}_J (x)=\frac{1}{\sqrt{2}}\,
g_\lambda(\eta)\,\Phi^*_J({\bf x}) \ ,
\ \  \ \
\tilde{\phi}^{(-)}_{\bar{J}} (x)=\frac{1}{\sqrt{2}}\,
g^*_\lambda(\eta)\,\Phi_{\bar{J}}({\bf x}) \ ,
\label{12}
\end{equation}
    $\Phi_J({\bf x}) $ is orthonormal eigenfunctions of $\Delta_{N-1}$
operator,   and
$\bar{J} $ is a set of quantum numbers of the function complex conjugated
to the function $\Phi_J $. \
In  ($N-1$)-dimensional spherical coordinates for
$J=\{\lambda \,, l\,,\ldots \, , m \} \ $    we have
$\bar{J}\!~=~\!\{\lambda \,, l\,,\ldots \, , - m \}  $.

    Substituting expansion~(\ref{11}) in the expression for
conserved charge
\begin{equation}
Q=i \int \limits_{\Sigma}\left(\tilde{\phi}^* \partial_0\tilde{\phi}
- (\partial_0\tilde{\phi}^* )\,\tilde{\phi}\,\right)
\sqrt{\gamma}\,d^{N-1}x \,,
\label{13}
\end{equation}
    where $\gamma=\mbox{det}(\gamma_{\alpha \beta})$,
    $\Sigma $ is a space-like hypersurface
$\eta\!=\!\mbox{const}$, and imposing the normalization condition
\begin{equation}
g_\lambda\,g_\lambda^{* \prime } -
g_\lambda^\prime\,g_\lambda^* = -2 i \ ,
\label{norm}
\end{equation}
    we obtain
\begin{equation}
Q = \int\! d \mu (J) \, \left(\stackrel{*}{a}\!{\!}^{(+)}_J a^{(-)}_J -
\stackrel{*}{a}\!{\!}^{(-)}_{\bar{J}} a^{(+)}_{\bar{J}} \right) \ .
\label{15}
\end{equation}

    The metrical Hamiltonian is expressed in terms of the metrical
energy-momentum tensor~(\ref{5}) by~\cite{GMM}:
\begin{eqnarray}
H(\eta)=\int \limits_\Sigma \zeta^i \, T_{ik}(x)\,d \sigma^k = \!\!
\int \limits_{\eta={\rm const}}\!\!\! \zeta^0 \,T_{00}(x) \,
g^{00}\sqrt{|g|} \ d^{N-1}x & = &       \nonumber   \\
= a^{N-2}(\eta) \!\! \int \limits_{\eta={\rm const}}\!\!\!T_{00}(x) \,
\sqrt{\gamma}\ d^{N-1}x \, ,\phantom{xxxxxxx}   & &
\label{16}
\end{eqnarray}
    where $ (\zeta^i) = (1, 0 , \ldots, 0) $ is the time-like conformal
Killing vector.

    The quantization is realized by the commutation relations
\begin{equation}
\left[a_J^{(-)}, \ \stackrel{*}{a}\!{\!}_{J'}^{(+)}\right]=
\left[\stackrel{*}{a}\!{\!}_J^{(-)}, \ a_{J'}^{(+)}\right]=\delta_{JJ'}  \ ,
\ \ \ \left[a_J^{(\pm)}, \ a_{J'}^{(\pm)}\right]=
\left[\stackrel{*}{a}\!{\!}_J^{(\pm)}, \ \stackrel{*}{a}\!{\!}_{J'}^{(\pm)}
\right]=0
\,.  \label{17}
\end{equation}
    The Hamilton operator (\ref{16}) can be written, through the
$ a_J^{(\pm)} , \ \stackrel{*}{a}\!{\!}_J^{(\pm)} $ operators in the form
\begin{eqnarray}
H(\eta)&=&\int d\mu(J)  \biggl\{ E_J(\eta) \,
\left(\stackrel{*}{a}\!{\!}^{(+)}_J a^{(-)}_J +
\stackrel{*}{a}\!{\!}^{(-)}_{\bar{J}} a^{(+)}_{\bar{J}} \right) +
  \nonumber \\
&+& F_J(\eta) \, \stackrel{*}{a}\!{\!}^{(+)}_J a^{(+)}_{\bar{J}} +
F^*_J(\eta)\, \stackrel{*}{a}\!{\!}^{(-)}_{\bar{J}} a^{(-)}_J  \biggr\} \,,
\label{H}
\end{eqnarray}
    where
\begin{equation}
E_J(\eta)=\frac{1}{2}\left\{\,|g_\lambda'|^2+
D_\lambda(\eta)\,|g_\lambda|^2 -
Q(\eta)\,(|g_\lambda|^2)' \, \right\}  \,,
\label{EJ}
\end{equation}
\begin{equation}
F_J(\eta)=\frac{(-1)^m}{2}\left\{{g_\lambda'}^2+
D_\lambda(\eta)\,g_\lambda^2 -
Q(\eta)\,(g_\lambda^2)'   \, \right\}  \,,
\label{FJ}
\end{equation}
\begin{equation}
D_\lambda(\eta)=m^2 a^2+\lambda^2+\Delta\xi\,(N-1)\,(N-2)\,(c^2-K) \,,
\label{21}
\end{equation}
\begin{equation}
Q(\eta)=\Delta \xi \, 2 \,(N-1) \, c    \,,
\label{22}
\end{equation}
   and  $c=a'(\eta)/ a(\eta)$ .
    The Hamilton operator corresponding to canonical energy-momentum
tensor~(\ref{4}) has the form (\ref{H}) with
(\ref{21}) and (\ref{22}) replaced by
\begin{eqnarray}
D_\lambda(\eta)&=&m^2 a^2+\lambda^2+(N-1)\,(N-2)\,\left(\,(\xi+\xi_c)\,
c^2 - \Delta \xi\,K \, \right) + \nonumber  \\*
&& \phantom{xxxxxxxx} +2\,\xi\,(N-1)\,c'  \,,
\label{34}
\end{eqnarray}
\begin{equation}
Q(\eta)=c\, (N-2)/\,2    \,.
\label{35}
\end{equation}

\section{The diagonalization of the metrical and\\
         canonical Hamilton operators}

\hspace{\parindent}
    The  Hamilton operator~(\ref{H}) is diagonal at time moment
$\eta_0 $ in the operators
$ \stackrel{*}{a}\!{\!}^{(\pm)}_J, \  a^{(\pm)}_J $,
which in this case are the creation and annihilation operators of particles
and antiparticles under  $F_J(\eta_0)=0$.
    Utilizing (\ref{FJ}) -- (\ref{35}) it may be shown that this condition
is consistent with normalization~(\ref{norm}) only when
$p^2_\lambda(\eta_0)>0 $,
  where for the metrical Hamilton operator
\begin{equation}
p_\lambda(\eta)=\sqrt{m^2a^2(\eta) + \lambda^2 + 4\Delta\xi \,
(N-1)^2\,(\xi\, c^2 -\xi_c \, K \,)} \,
\label{23}
\end{equation}
   and for the canonical Hamilton operator
\begin{equation}
p_{\lambda}(\eta) = p_{can,\lambda}(\eta) =
\sqrt{(m^2+\xi\,R)\,a^2(\eta) + \lambda^2 - ((N-2)/\,2)^2\,K} \,.
\label{36}
\end{equation}

    The requirement for the metrical Hamilton operator to be diagonal
at the instant $\eta_0$, i.e., $F_J(\eta_0)=0$,
and the normalization condition lead to the initial conditions on the
functions   $g_\lambda(\eta_0)$:
\begin{equation}
g_\lambda'(\eta_0)=( 2 \Delta\xi (N-1)\, c +
i p_\lambda(\eta_0))\, g_\lambda(\eta_0) \,, \ \
|g_\lambda(\eta_0)|= 1/\sqrt{p_\lambda(\eta_0) } \,.
\label{24}
\end{equation}
    For the canonical Hamilton operator the initial conditions are
\begin{equation}
g_\lambda'(\eta_0)=((N-2)\,c/\,2 +
i\, p_{can,\lambda}(\eta_0))\, g_\lambda(\eta_0) \,, \ \
|g_\lambda(\eta_0)|= 1/\sqrt{p_{can,\lambda}(\eta_0) } \,.
\label{37}
\end{equation}
    The state of vacuum $|\,0\!>$ corresponding to (\ref{24}), (\ref{37})
is defined in the standard form
\begin{equation}
a^{(-)}_J|\,0\!> \,=\, \stackrel{*}{a}\!{\!}^{(-)}_J|\,0\!  >\, = 0 \,.
\label{25}
\end{equation}

    For arbitrary time moment $\eta $ we diagonalize the Hamilton
operator in terms of
$ \stackrel{*}{b}\!{\!}^{(\pm)}_J, \  b^{(\pm)}_J $ operators which
    connected with
$ \stackrel{*}{a}\!{\!}^{(\pm)}_J, \  a^{(\pm)}_J $,
    by the time-dependent Bogolyubov transformations:
\begin{equation}
\left\{  \begin{array}{c}
a_J^{(-)}=\alpha^*_J(\eta) \,b^{(-)}_J(\eta)-
(-1)^m \beta_J(\eta)\, b^{(+)}_{\bar{J}}(\eta) \,,  \\[3mm]
\stackrel{*}{a}\!{\!}_J^{(-)}=\alpha^*_J(\eta) \,
\stackrel{*}{b}\!{\!}^{(-)}_J\!(\eta)-
(-1)^m \beta_J(\eta) \,\stackrel{*}{b}\!{\!}^{(+)}_{\bar{J}}\!(\eta) \,,
\end{array} \right.
\label{26}
\end{equation}
  where $\alpha_J(\eta), \ \beta_J(\eta) $ are the functions satisfying
the initial conditions         \\ $|\alpha_J(\eta_0)|=1, \
\beta_J(\eta_0)=0 $ and the identity
\begin{equation}
|\alpha_J(\eta)|^2-|\beta_J(\eta)|^2=1       \,.
\label{27}
\end{equation}
(In homogeneous and isotropic space
$\alpha_J =\alpha_\lambda , \ \beta_J=\beta_\lambda $ \cite{GMM}).

    The substitution of the decomposition (\ref{26}) in (\ref{H})
gives, if we demand coefficients before nondiagonal terms
$\stackrel{*}{b}\!{\!}^{(\pm)}_J b^{(\pm)}_J $ equal to 0,
the  equation
\begin{equation}
2 (-1)^{m+1} \alpha_J \beta_J E_J+ F_J \alpha_J^2+F_J^* \beta_J^2=0 \,.
\label{28}
\end{equation}
    It can be shown that the condition (\ref{28}) is consistent with
the normalization~(\ref{norm}), only if $p^2_\lambda(\eta)>0$.
    In that case
\begin{equation}
|\beta_J|^2=E_J/\,(2p_\lambda)-1/2=|F_J|^2/\,(2 p_\lambda \,
(E_J+p_\lambda ))    \,.
\label{29}
\end{equation}
    In obtaining (\ref{29}) we take into account, the result
that can be checked,
\begin{equation}
E_J^2- |F_J|^2 = p^2_\lambda(\eta) \!\cdot\!
\left[  - ( g_\lambda\,g_\lambda^{* \prime } -
g_\lambda^\prime\,g_\lambda^*\, )^2 /\,4\,\right] \,.
\label{30}
\end{equation}
    (The multiplier in square brackets equals 1 under the normalization
condition~(\ref{norm})).

    In the case of (\ref{28}) and $ p^2_\lambda(\eta)>0 $,
the Hamilton operator (\ref{H}) takes the form
\begin{equation}
H(\eta) =\int d\mu(J) \,p_\lambda(\eta) \,
\left(\,\stackrel{*}{b}\!{\!}^{(+)}_J b^{(-)}_J +
\stackrel{*}{b}\!{\!}^{(-)}_{\bar{J}} b^{(+)}_{\bar{J}}\, \right) \,.
\label{31}
\end{equation}
    So $p_\lambda(\eta) $ has the meaning of energy of quasiparticles
corresponding to the diagonal form of the metrical Hamilton operator
(and $p_{can, \lambda}(\eta) $ for the canonical Hamilton operator).
    For the 4-dimensional space-time the equation (\ref{23})
corresponds to energy values obtained in \cite{Diag} and~\cite{CF86}\,.

    The quasiparticle energy $p_\lambda(\eta) $ differs from the oscillator
frequency $\Omega(\eta) $ of the wave equation for nonconformal field,
   and this leads to a series of difficulties.
Thus the conditions $p^2_\lambda(\eta)>0 $ and $\ \Omega^2(\eta)>0
$ may be in contradiction for a nonconformal field in some cases.
For example, in the case of quasi-Euclidean space ($K=0 $) and
zero-mass field the condition $p^2_\lambda(\eta)>0 $ (with
arbitrary~$\lambda$) for the metrical case reduces to $\ \xi \in
[\, 0 , \,\xi_c \,] $; but if $\xi<\xi_c \,, \ m=0 $ and  $R>0 $
for low $\lambda $ then we have $\ \Omega^2(\eta)< 0 $.

    It should be noted that for $p^2_\lambda(\eta)<0 $ the condition
of diagonalization reduces to the vanishing of norm, energy and charge of
the state with $\phi(x)\ne 0 $,
and this situation does not have any physical foundation.

    The vacuum state defined by the equations
\begin{equation}
b^{(-)}_J|\,0_\eta\!> \,=\, \stackrel{*}{b}\!{\!}^{(-)}_J|\,0_\eta\!>\,
= 0 \,,
\label{32}
\end{equation}
   depends on time in the nonstationary metric.
Under the initial conditions (\ref{24}), (\ref{37})   we have
$ b^{(\pm)}_J(\eta_0)=a^{(\pm)}_J $ and $|\,0_{\eta_0}\!>=|\,0\!> $.
In the Heisenberg representation, the state $|\,0\!>$, which is vacuum
at the instant $\eta_0$, is no longer a vacuum for $\eta\ne\eta_0$.
It contains $|\beta_J(\eta)|^2$ quasiparticle pairs corresponding to
the operators
$\stackrel{*}{b}\!{\!}^{(\pm)}_J\,,   b^{(\pm)}_J$
in every mode~\cite{GMM}.
    The number of the created pairs of quasiparticles in the unit of space
volume (for $N=4$) is \cite{GMM}
\begin{equation}
n(\eta)=\frac{1}{2 \pi^2 a^3(\eta)}
\int d\mu(J)\,|\beta_\lambda(\eta)|^2 \,.
\label{33}
\end{equation}
   For asymptotic solutions of equation (\ref{8}) (see \cite{Fed}),
normalized according to (\ref{norm}),
we can obtain from (\ref{EJ})---(\ref{35})
that $E_J \sim \lambda $ and for nonstationary metrics
$|F_J(\eta)| \sim |Q(\eta)| $ in  $\lambda \to \infty $.
Therefore, according to (\ref{29}), this is corrected with the substitution
of $p_\lambda \rightarrow p_{can,\lambda}$, and we have
$|\beta_\lambda|^2 \sim \lambda^{-2} $.
    Consequently, the density of created quasiparticles, proportional to
the integral in (\ref{33}), is infinite.

    So, in the diagonalization procedure, both for the metrical and the
canonical Hamilton operators in nonconformal scalar fields, there is
a problem of infinite density of quasiparticles created in
the nonstationary metrics.
     In both cases the energies of corresponding quasiparticles differ
from the oscillator frequency of the wave equation.
    It is shown below that these difficulties are absent in the case
of the Hamilton operator corresponding to the modified energy-momentum
tensor.

\section{Modified energy-momentum tensor and\\
         modified Hamilton operator}

    Let us consider the modified energy-momentum tensor
\begin{equation}
T^{\,mod}_{ik}=T_{ik}^{can}-2 \xi_c\, [R_{ik}+\nabla_i\nabla_k-g_{ik}
\nabla_j\nabla^j ] \, \phi^* \phi \, .
\label{Tmod}
\end{equation}
    From the definition (\ref{Tmod}) it is clear that for conformal scalar
fields (i.e. if $\xi=\xi_c $) \  $T^{\,mod}_{ik} $ coincides with
the metrical energy-momentum tensor~(\ref{5}).
    The structure of the Hamiltonian constructed by $T^{\,mod}_{ik} $
similarly to (\ref{16}) is
\begin{eqnarray}
H^{mod}(\eta) &=& \! \int h(x) \, d^{N-1}x =
\int d^{N-1}x\,\sqrt{\gamma} \, \biggl\{
\tilde{\phi}^{* \prime} \tilde{\phi}'
+\gamma^{\alpha \beta}\partial_\alpha\tilde{\phi}^*
\partial_\beta\tilde{\phi}+  \nonumber     \\
&+& \! \Bigl[\, m^2a^2-\Delta \xi\, a^2 R +
\Bigl((N-2)/\,2\Bigr)^2 K\,\Bigr]\,
 \tilde{\phi}^* \tilde{\phi} \, \biggr\} \,.
\label{39}
\end{eqnarray}

    We show that the modified Hamiltonian (\ref{39}) can be
obtained in homogeneous isotropic space as canonical under the certain
choice of variables describing scalar field.
    If we add $N$-divergence $({\partial J^i}/{\partial x^i})$,
to the Lagrangian density~(\ref{3}),
where in the $(\eta, {\bf x})$ system of coordinates the $ N$-vector
$\ (J^i)=(\sqrt{\gamma}\,c\,\tilde{\phi}^*\, \tilde{\phi}\,(N-2)/2, \, 0,
\, \ldots \,, 0) $, \
   the movement equations (\ref{1}) are invariant under this addition.
Choosing
$\tilde{\phi}(x)= a^{(N-2)/\,2}(\eta)\phi(x) $
and $\tilde{\phi}^*(x)$, i.e., the variables in terms of which
the equation~(\ref{1}) has the form~(\ref{7}), for the field's coordinates
   and using the Lagrangian density
$ L^{\Delta}(x)=L(x)+({\partial J^i}/{\partial x^i})$, \
we obtain that the Hamiltonian density
$ \tilde{\phi}'\,(\partial L^{\Delta})/(\partial \tilde{\phi}')+
\tilde{\phi}^{* \prime}\,(\partial L^{\Delta})/
(\partial \tilde{\phi}^{* \prime})-L^{\Delta}(x) \ $
is equal to $h(x) $, from (\ref{39}).
    This is why the Hamiltonian (\ref{39}) is a canonical one for the scalar
field, if $\tilde{\phi}(x) $ and $\ \tilde{\phi}^*(x) $ are chosen as
the field's variables.

    The modified Hamilton operator can be written in form (\ref{H}),
but in that case $ Q(\eta)=0 $ and  $\ D_\lambda(\eta)=\Omega^2(\eta) $;
   under its diagonalization by
$\stackrel{*}{b}\!{\!}^{(\pm)}_J\,,   b^{(\pm)}_J$,
operators we obtain (\ref{31}) with the change
$p_\lambda \rightarrow \Omega $.
The oscillator frequencies $\Omega(\eta) $ then coincide with the energy
of corresponding particles.
    The initial conditions  for $g_\lambda(\eta) $, corresponding
to the diagonal form in the time moment $\eta_0 $ with operators
$\stackrel{*}{a}\!{\!}^{(\pm)}_J\,,   a^{(\pm)}_J$ (\ref{17}), are
\begin{equation}
g_\lambda'(\eta_0)=i\, \Omega(\eta_0)\, g_\lambda(\eta_0) \,, \ \ \
|g_\lambda(\eta_0)|= 1/\sqrt{\Omega(\eta_0) } \,.
\label{40}
\end{equation}
   They coincide with the initial conditions used in \cite{BLM} if
$\arg g_\lambda(\eta_0)=0 $ is fixed.
  In the case of radiation dominated background ($R=0$) they coincides with
conditions used in \cite{Pv, MMSH}.

    We show that the density of the particles corresponding to
the diagonal form of $H^{\,mod} $ and created in the nonstationary metric
is finite.
    For this, we find the asymptotic behavior of the
functions $|\beta_\lambda(\eta)|^2 $ as $\lambda \to \infty $.
    The functions $\beta_\lambda(\eta) $ and $\ \alpha_\lambda(\eta) $
that are the solutions of~(\ref{28}) and satisfy  identity~(\ref{27})
can be represented as
\begin{equation}
\beta_\lambda(\eta)=\frac{i}{2} \frac{e^{i\,\Theta(\eta_0, \eta)}}
{\sqrt{\Omega}}\, \biggl( g'(\eta)-i\,\Omega\, g(\eta)\biggr)  \,,
\label{41}
\end{equation}
\begin{equation}
\alpha_\lambda(\eta)=\frac{i}{2} \frac{e^{i\,\Theta(\eta_0, \eta)}}
{\sqrt{\Omega}}\, \biggl( g^{* \prime}(\eta)-i\,\Omega\, g^*(\eta)
\biggr)  \,,
\label{42}
\end{equation}
    where
$ \Theta(\eta_1, \eta_2) = \int \limits_{\eta_1}^{\eta_2}
\Omega(\eta)\,d\eta $.
  \  In consequence of (\ref{41}), (\ref{42}) and equation (\ref{8})
the functions
$s_\lambda(\eta)=|\beta_\lambda(\eta)|^2 $ and
$\ f_\lambda(\eta)=2\,\alpha_\lambda(\eta)\, \beta_\lambda(\eta)
\exp[-2i\,\Theta(\eta_0,\eta)] $
    satisfy  the system of equations:
\begin{equation}
\left\{  \begin{array}{l}
s_\lambda'(\eta)={\displaystyle \frac{\Omega'}{2\,\Omega}}\,
{\rm Re} f_\lambda(\eta)    \,,  \\[3mm]
f_\lambda'(\eta)+2\,i\,\Omega\,f_\lambda(\eta)=
{\displaystyle \frac{\Omega'}{\Omega}} \,(1+2 s_\lambda(\eta)) \,.
\end{array} \right.
\label{43}
\end{equation}
    Taking into account the initial condition
$ s_\lambda(\eta_0)=f_\lambda(\eta_0)=0 $
(as $\beta_\lambda(\eta_0)=0 $) we write the system of differential
equations (\ref{43}) in the equivalent form of the system of Volterra
integral equations
\begin{equation}
f_\lambda(\eta)=\int_{\eta_0}^\eta w(\eta_1)\,
(1+2 s_\lambda(\eta_1))\, \exp[-2\,i\,\Theta(\eta_1,\eta)]\,d\eta_1  \,,
\label{44}
\end{equation}
\begin{equation}
s_\lambda(\eta)=\frac{1}{2}\,\int_{\eta_0}^\eta d\eta_1 \,
w(\eta_1)\, \int_{\eta_0}^{\eta_1} d\eta_2 \,w(\eta_2)\,
(1+2 s_\lambda(\eta_2)) \cos[2\,\Theta(\eta_2,\eta_1)]  \,,
\label{45}
\end{equation}
    where $w(\eta)=\Omega'(\eta)/\,\Omega(\eta) $.
    To find the asymptotic behavior of $ s_\lambda(\eta) $, we restrict
our consideration to the first iteration of integral equation~(\ref{45})
and,taking into account that
$ \Theta(\eta_2, \eta_1 )\to \lambda (\eta_1-\eta_2) $
as $\lambda \to \infty $, represent~(\ref{45}) as
    \begin{equation}
s_\lambda(\eta) \approx \frac{1}{4}\,\left|\,\int_{\eta_0}^\eta
w(\eta_1)\,\exp( 2\,i\,\lambda\,\eta_1 )\,d\eta_1 \right|^2.
\label{46}
\end{equation}
    Consequently, we have $s_\lambda \sim \lambda^{-6} $,
and the integral in~(\ref{33}) is therefore convergent.
    Thus in this case the density of created particles is finite for
4-dimensional space-time.
    In the case of finite volume space ($ K=+1 $) the total number
of created particles is finite also, the Bogolyubov
transformations realized by Hilbert-Schmidt operators, and the
representations of commutation relations for operators
$\stackrel{*}{b}\!{\!}^{(\pm)}_J(\eta)\,,   b^{(\pm)}_J(\eta) $
are unitarily equivalent for all time.

    In the presented work the metrical, canonical and introduced modified
Hamilton operators are investigated.
   It is shown that the density of particles created in nonstationary
homogeneous isotropic space metric is finite only in the case of
modified Hamiltonian (\ref{39}) and the energies of such particles are
equal to the oscillator frequency.

    The modified energy-momentum tensor (\ref{Tmod}),
introduced above, coincides
with the metrical one for a conformal scalar field.
    In homogeneous isotropic space $T_{ik}^{\,mod} $ results in the modified
Hamiltonian (\ref{39}) which can be obtained as well as canonical under
the special choice of field's variables.

    It can be seen that considering a line combination of metrical (\ref{5})
and canonical (\ref{4}) tensors we can certainly obtain the modified tensor
(\ref{Tmod}) if the quasiparticles' energy coincides with oscillator
frequency.
     It should be stressed that the metrical energy-momentum tensor can not
be changed to $T_{ik}^{\,mod} $ in the right-hand sides of
Einstein's equations because $T_{ik}^{\,mod} $ is not covariant
conservation.
    However under the corpuscular interpretation of the nonconformal
scalar field and when the diagonalization procedure is used,
the modified Hamilton operator constructed by $T_{ik}^{\,mod} $
is preferable in comparison with the metrical Hamilton operator.\\[7mm]
 {\bf Acknowledgments.}
The author is grateful to Prof. A.A.Grib for helpful discussions.

\end{document}